\documentclass[fleqn,usenatbib]{mnras}

\usepackage{newtxtext,newtxmath}

\usepackage[T1]{fontenc}
\usepackage{ae,aecompl}

\usepackage{graphicx}	
\usepackage{amsmath}	
\usepackage{amssymb}	
\usepackage{siunitx}

\newcommand{\T}{\mathcal{T}}
\newcommand{\Rp}{R_\mathrm{P}}

\title[An empirical transit spectrum of Earth]{An empirical infrared transit spectrum of Earth: \\opacity windows and biosignatures}

\author[Evelyn J. R. Macdonald and Nicolas B. Cowan]{
Evelyn J. R. Macdonald,$^{1,2,3}$\thanks{E-mail: evelyn.macdonald@mail.mcgill.ca} and
Nicolas B. Cowan$^{1,2,3,4}$\thanks{E-mail: nicolas.cowan@mcgill.ca}
\\
 $^{1}$Department of Physics, McGill University, 3600 rue University, Montr\'eal, QC H3A 2T8, Canada\\
 $^{2}$McGill Space Institute, 3550 rue University, Montr\'eal, QC H3A 2A7, Canada\\
 $^{3}$Institut de recherche sur les exoplan\`etes, Universit\'e de Montr\'eal, C.P. 6128, Succ. Centre-ville, Montr\'eal, QC H3C 3J7, Canada\\
 $^{4}$Department of Earth \& Planetary Sciences, McGill University, 3450 rue University, Montr\'eal, QC H3A 0E8, Canada
}

\date{Accepted XXX. Received YYY; in original form ZZZ}

\pubyear{2018}

\begin{document}
\label{firstpage}
\pagerange{\pageref{firstpage}--\pageref{lastpage}}
\maketitle

\begin{abstract}
The Atmospheric Chemistry Experiment's Fourier Transform Spectrometer on the \textit{SCISAT} satellite has been measuring infrared transmission spectra of Earth during Solar occultations since 2004. We use these data to build an infrared transit spectrum of Earth. Regions of low atmospheric opacity, known as windows, are of particular interest, as they permit observations of the planet's lower atmosphere. Even in the absence of clouds or refraction, imperfect transmittance leads to a minimum effective thickness of $h_{\rm min} \approx 4$~km in the 10--$\SI{12}{\micro m}$ opacity window at a spectral resolution of $R=10^3$. Nonetheless, at $R=10^5$, the maximum transmittance at the surface is around $\SI{70}{\%}$. In principle, one can probe the troposphere of an Earth-like planet via high-dispersion transit spectroscopy in the mid-infrared; in practice aerosols and/or refraction likely make this impossible. We simulate the transit spectrum of an Earth-like planet in the TRAPPIST-1 system. We find that a long-term near-infrared campaign with \textit{JWST} could readily detect CO$_2$, establishing the presence of an atmosphere. A mid-IR campaign or longer NIR campaign would be more challenging, but in principle could detect H$_2$O and the biosignatures O$_3$ and CH$_4$.
\end{abstract}


\section{Introduction}

Transit spectroscopy is an observational method in which the flux of a star is measured as a function of wavelength during a planet's transit. The partial obstruction of the star by the planet causes a decrease in flux proportional to the planet's cross-sectional area. Absorption and scattering of photons by the planet's atmosphere cause an additional, wavelength-dependent flux decrease \citep[][]{Kreidberg2017}. A rocky planet's apparent radius is equal to the radius of the solid planet plus a wavelength-dependent term representing the contribution of its atmosphere to its transit depth; this is the atmosphere's effective thickness. 

The transit spectrum of a planet is an alternating sequence of low-opacity windows, in which one can potentially probe the near-surface environment, and absorption features that probe higher layers of the atmosphere. A planet's atmospheric composition can be inferred by attributing wavelengths of large apparent radius to specific molecules \citep[][]{Madhusudhan2018}. 
The presence of certain molecules in a planet's spectrum can be attributed to life on the planet; these are biosignatures (see \citealt{kiang_review_2018} for review). In particular, methane and oxygen are in chemical disequilibrium on Earth due to their continuous production by living organisms \citep{sagan_biosignatures_1993, segura_biosignatures_2005}. The detection of both in an exoplanet's atmosphere would strongly suggest that the planet was inhabited \citep{kaltenegger_review_2017}.

Previous studies have used Earth's atmosphere to model spectra of other planets. \citet{kaltenegger_transits_2009} developed an Earth atmosphere model---validated against ATMOS 3 Solar occultation data---and used it to produce an effective thickness spectrum of our planet \citep[see also][]{ehrenreich_2006, rauer_biosignatures_2011, howe_2012, hedelt_2013, vasquez_2013}. \cite{palle_2009} used optical--NIR observations of the moon during a lunar eclipse to study Earth's atmosphere. \citet{schreier_transmission_2018} made effective thickness spectra using the same data as we use in the present paper.
Others have studied the prospects for characterizing transiting terrestrial exoplanets with the \textit{James Webb Space Telescope} \citep[\textit{JWST};][]{deming_2009, misra_2014, Beichman2014,cowan_characterizing_2015}. 

In this paper, we use satellite data to build an empirical infrared transit spectrum of Earth \citep[for further examples of Earth-as-an-exoplanet, see][]{RobinsonReinhard2018}. We describe the data in Section~\ref{sec:data} and use them to construct transit spectra in Section~\ref{sec:methods}. We then consider differences in host star type, accounting for refraction and photon noise in Section~\ref{realistic_data}. We discuss our findings in Section~\ref{sec:conclusion}.

\section{Solar Occultation Data}\label{sec:data}

\begin{figure}
\includegraphics[width=\columnwidth]{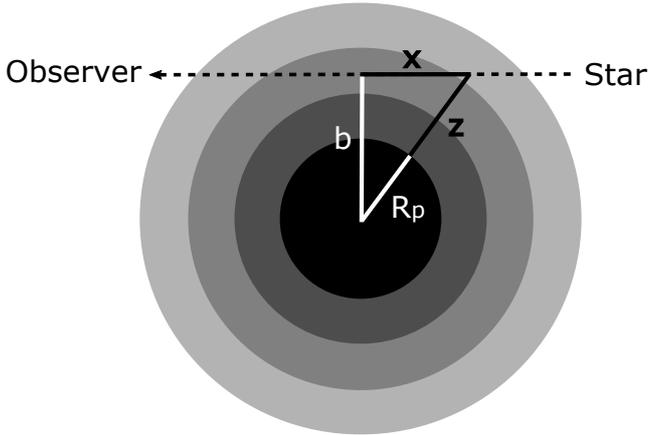}
\caption{Light (dotted line) passing through the atmosphere of a planet of radius $\Rp$, along a particular chord in the $\hat{x}$ direction. The impact parameter of the chord is $b$ and $z$ is the altitude of a parcel of gas above the surface.}
\label{fig:geometry}
\end{figure}

In this work, we construct an empirical transit spectrum of Earth using Solar occultation spectra. The geometry of Solar occultation is shown in Fig. \ref{fig:geometry}. This geometry is strikingly similar to that of an exoplanet in transit, except that in the latter case one simultaneously probes all impact parameters and therefore all atmospheric layers. 

We use data from the \textit{Atmospheric Chemistry Experiment Fourier Transform Spectrometer (ACE-FTS)}\footnote{Data taken from \url{http://www.ace.uwaterloo.ca/atlas.php}} on the Canadian low-Earth orbit satellite \textit{SCISAT}\footnote{\url{http://www.asc-csa.gc.ca/eng/satellites/scisat/}}. \textit{ACE-FTS} measures the atmosphere's transmittance, the fraction of incident sunlight which reaches the instrument after travelling through the atmosphere. We describe the data acquisition below; more details can be found in \cite{bernath_2001}, \cite{hughes_ace_2014}, and \cite{bernath_ace_2017}.

The \textit{ACE} mission is designed to study trace atmospheric constituents; consequently, the resolution of the data is high. \textit{SCISAT}'s orbit is circular, with an inclination of $74^\circ$ with respect to the equator and an altitude of $\SI{650}{km}$. The instrument takes 30 measurements per day except during two three-week periods, one in June and one in December, during which the geometry prevents it from recording spectra.

During a sunrise or sunset, \textit{ACE-FTS} records a spectrum of its entire wavenumber range of $100$-$\SI{5000}{cm^{-1}}$ (2--100~$\mu$m) every two seconds. Because the duration of a sunrise or sunset depends on the time of year, the number of spectra recorded per occultation ranges from 27 to 65. During a sunset, the instrument starts at a high altitude, where it can take spectra of the Sun without interference by the atmosphere. As the satellite orbits, its impact parameter decreases, allowing it to take spectra repeatedly until the Sun disappears from its field of view, either due to reaching an impact parameter below $\SI{5}{km}$ or due to obstruction by clouds. Spectra affected by clouds are removed, so those used in our study are cloud-free. \footnote{Visible and near-infrared filtered imagers aboard \textit{SCISAT} can be used to study clouds using solar extinction profiles \citep{Dodion}.} 
The sampling at low altitudes is compressed due to the effects of refraction. 

The Polar vortex is identified using derived meteorological products and removed from the data. The data are then divided into 4-km bins and averaged, such that each averaged spectrum is composed of around 800 spectra. The signal-to-noise ratio is wavenumber-dependent in the raw spectra, and increases as the square root of the number of spectra included in the average. Its continuum value in the published data is around 8000.

\begin{figure}
    \centering
    \includegraphics[width=\columnwidth]{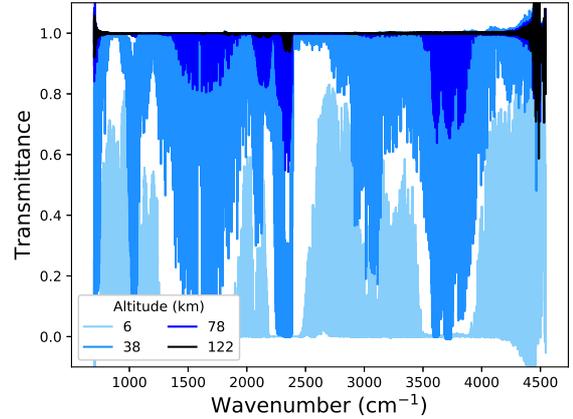}
    \caption{Sample raw data from ACE-FTS: transmittance as a function of wavenumber for four sample 4~km altitude bins in the mid-latitude summer dataset. Bins are labelled by their mean altitude. The input data plotted here are averages of many Solar occultations free of stratospheric clouds and therefore represent a cloud-free time average. As expected, the transmittance is essentially unity high up but drops with altitude, and exhibits strong wavelength-dependence due to molecular absorption lines.}
    \label{fig:rawdata}
\end{figure}

The transmittance is measured as a function of wavenumber, sampled every $\SI{0.0025}{cm^{-1}}$ for altitudes of 4--$\SI{128}{km}$ in $\SI{4}{km}$ bins. The average altitude of each bin is provided. There are five datasets corresponding to three latitude regions: Arctic summer and winter (60--$\SI{85}{^\circ N}$), mid-latitude summer and winter (30--$\SI{60}{^\circ N}$), and tropics ($\SI{30}{^\circ S}$--$\SI{30}{^\circ N}$). The data are provided for 100--$\SI{5000}{cm^{-1}}$ or 2--$\SI{100}{\micro m}$; we consider the 2.2--$\SI{14.3}{\micro m}$ range, as no spectral features could be identified elsewhere due to measurement noise. Fig. \ref{fig:rawdata} shows a sample of the raw data for the mid-latitude summer dataset. Note that these data are composed of averaged cloud-free Solar occultation spectra, as described above.

Transmittance should be between 0 and 1: to correct the $\sim$ 4\% percent of bad data, we set the transmittance to 0 if it is negative and to 1 if it is above 1. Bad data result from measurement or calibration error, which happens when the transmittance is near the edges of the acceptable range. Truncation is the most accurate way to correct for these errors. For a given wavelength, transmittances at every altitude are needed to calculate the effective thickness; therefore, correcting the bad data is necessary to avoid gaps in the spectrum.

We treat the Northern and Southern Hemispheres as identical, so each latitude range represents 1/3 of the planet's circumference. By averaging the summer and winter datasets for the Arctic, as well as for the mid-latitudes, we build year-round models for each of these regions. This process is unnecessary for the tropics due to the lack of seasonal variation. This seasonal averaging is justified because transit spectroscopy probes the day-night terminator, which is a great circle. If the planet is tidally locked, the terminator will be at the intersection of warm and cold climates; otherwise, non-tropical regions along the terminator are either at temperate climates, or equally split between winter and summer. We average our three year-round spectra to make a global year-round average. This can be thought of as a snapshot of the planet's overall climate, as it would be seen in transit.

\begin{figure*}
\includegraphics[width=\textwidth]{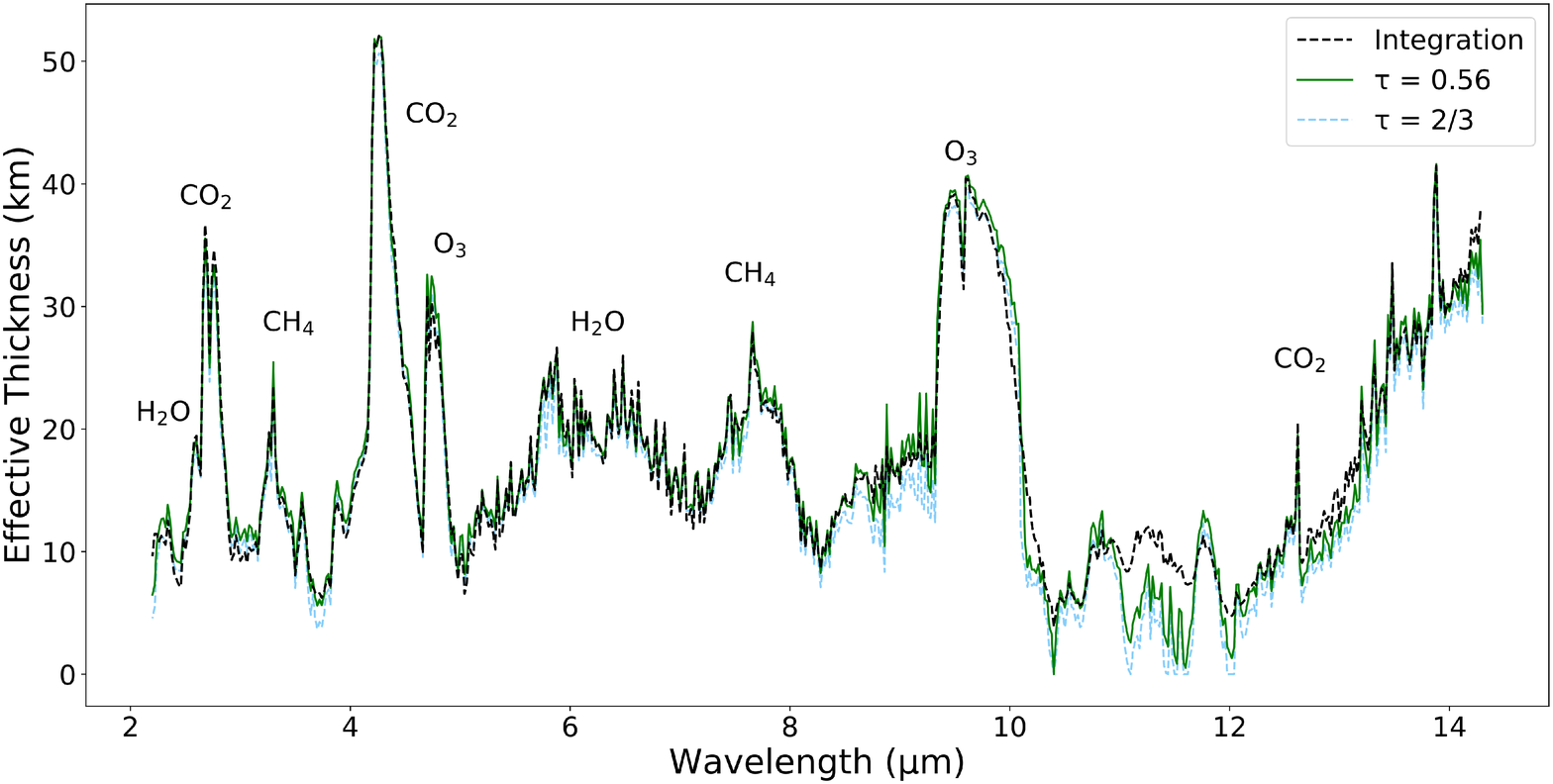}
\caption[]{Effective thickness spectra of Earth made with the integration method (black dotted line) and optical depth approximation, with $\tau=0.56$ (green) and $\tau=2/3$ (blue). These spectra represent globally-averaged spectra (Section \ref{sec:data}) and the spectral resolution has been reduced to $\SI{0.02}{\micro m}$. Molecules are labelled as in \protect\citet{kaltenegger_transits_2009} and \protect\citet{schreier_transmission_2018}. The simultaneous presence of O$_3$ and CH$_4$ is a biosignature \protect\citep{sagan_biosignatures_1993, segura_biosignatures_2005}. The $\tau=0.56$ and $\tau=2/3$ approximations generally agree with the integration approach but significantly underestimate the atmosphere's effective thickness in opacity windows (e.g., 10--12~$\mu$m).} 
\label{fig:tau}
\end{figure*}

\section{Synthetic Transit Spectra}\label{sec:methods}

\subsection{Integrations over Chords} \label{sec:td}
 
The transit depth, $D$, is the fractional decrease in flux due to the partial obstruction of the star by the planet. If we assume, for simplicity, that the planet is centred on the star, then the transit depth at a wavelength $\lambda$ is given by \citep{brown_2001, deWit_2013, betremieux_analytical_2017,heng_2017, Robinson_scattering_2017,jordan_2018}:

\begin{equation} \label{eq:integral}
D(\lambda) = \left(\frac{\Rp}{R_*}\right)^2 + \frac{2}{R_*^2}\int_{\Rp}^{R_*} b(1-\T(b,\lambda))db,
\end{equation}
where $\Rp$ and $R_*$ are the radii of the planet and star, $b$ is the impact parameter of a chord, and $\T$ is the atmospheric transmittance along that chord (see Figure~\ref{fig:geometry}). Conveniently, $\T(b,\lambda)$ is precisely the data product provided by \textit{ACE-FTS}.

Given a transit depth, the atmosphere's effective thickness, $h_\lambda$, can be computed by noting that transit depth is also given by:
\begin{equation} \label{eq:reff}
D = \frac{(\Rp+h_\lambda)^2}{R_*^2}.
\end{equation}

Combining Eqs.~\ref{eq:integral} and \ref{eq:reff}, solving for $h_\lambda$ using the quadratic formula, and recognizing that the effective thickness must be positive, one obtains 
\begin{equation}\label{h_eqn}
h_\lambda = -\Rp + \sqrt{\Rp^2+2\int_{\Rp}^{R_*} b(1-\T(b,\lambda))db}.
\end{equation}
In Appendix~\ref{Na} we derive an analytic approximation to this expression in the context of Earth's sodium D-lines.

Note that \cite{deWit_2013} and \cite{betremieux_analytical_2017} derive $h_\lambda$ in terms of the Euler-Mascheroni constant, $\gamma \approx 0.577$, of the form
\begin{equation}
h_\lambda = H\gamma + F_\lambda,    
\end{equation}
where $H$ is the atmospheric scale height and $F_\lambda$ is a wavelength-dependent atmospheric contribution. On the other hand, \cite{kaltenegger_transits_2009} and \cite{schreier_transmission_2018} calculate the effective thickness assuming a \emph{plane-parallel} atmosphere: $h_\lambda = \int_{\Rp}^{R_*}(1-\T(b))db$. This is only a valid approximation if the effective height of the atmosphere is much smaller than the planetary radius ($h_\lambda << \Rp$), which is not always true for exoplanets \citep[e.g.,][]{Ehrenreich2015}.

To build transit spectra, we approximate the integral of the impact parameter using the trapezoidal rule with $db \approx \SI{4}{km}$ based on the sampling of the data. To approximate the transmittance near the surface, we use linear extrapolation, except where the effective thickness is less than 12~km, as linear extrapolation overestimates the transmittance in transparent regions. We therefore estimate the near-surface transmittance by calculating the optical depth (Eq. \ref{eq:tau}) and fitting it to a decaying exponential in these regions. We set the upper limit of integration at 124--$\SI{128}{km}$ above the surface, based on availability of data. The black dotted curve of Fig. \ref{fig:tau} shows our effective thickness spectrum of Earth with a spectral resolution of $\Delta \lambda = \SI{0.02}{\micro m}$.\footnote{An alternative to binning is convolution with a Gaussian  \protect\citep{schreier_transmission_2018}. This approach results in a smoothed spectrum, and thus may be less characteristic of real data.}

\begin{figure}
\includegraphics[width=\columnwidth]{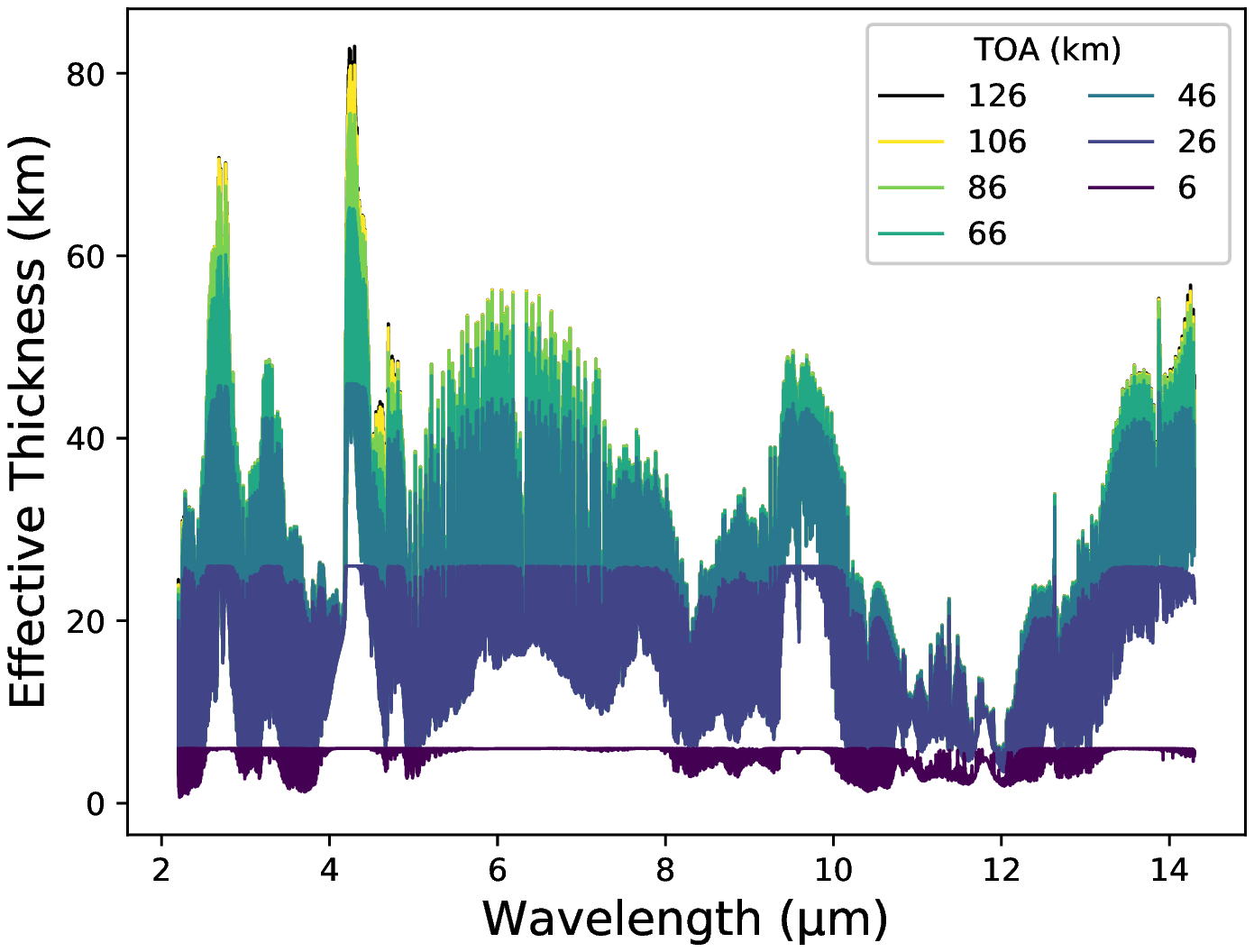}
\includegraphics[width = \columnwidth]{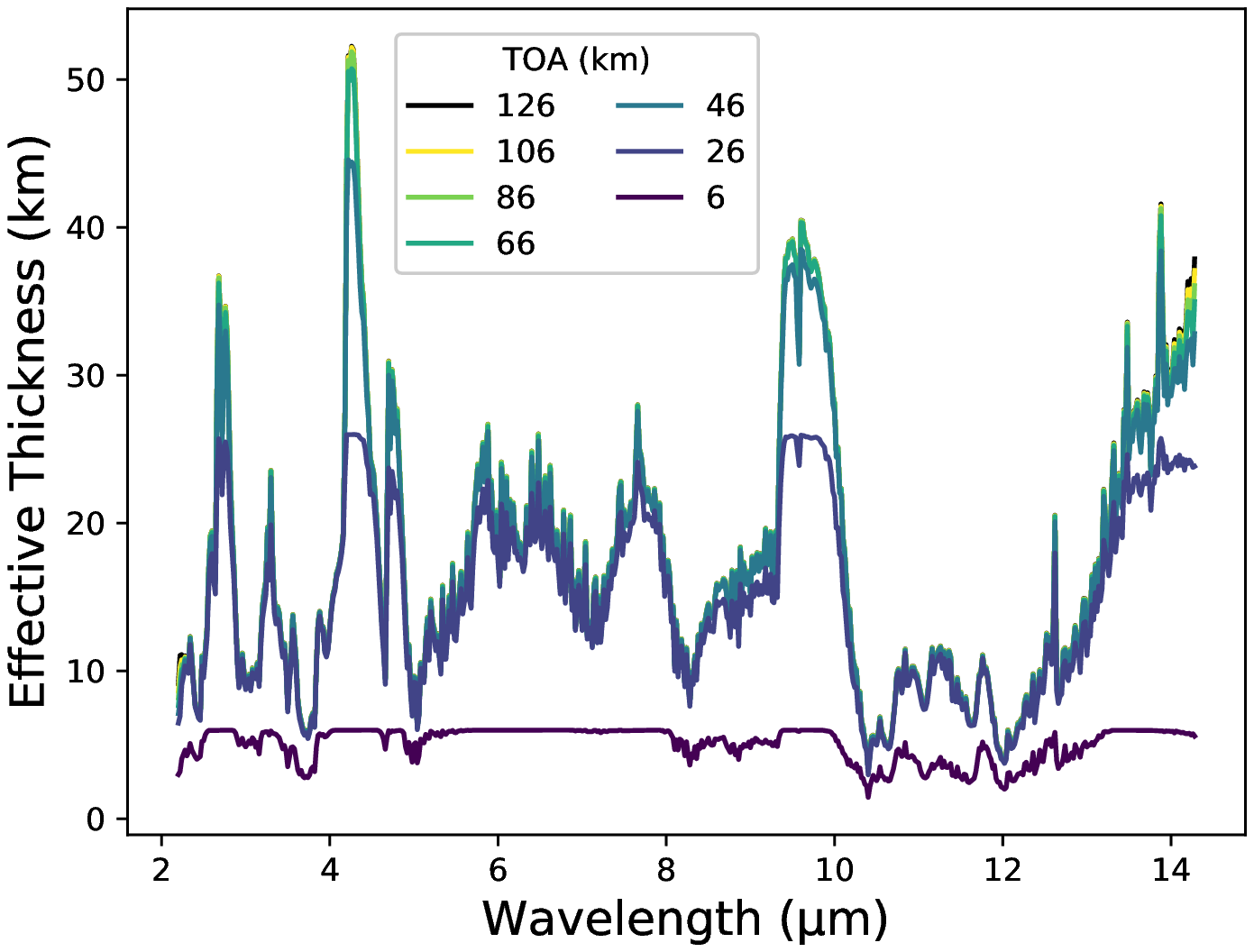}
\caption{Effective thickness spectrum of Earth for different  top-of-atmosphere (TOA) altitudes (the upper limit of integration in Eq.~\ref{eq:integral}). Only atmospheric layers near the surface contribute significantly to optically thin spectral features, whereas a large range of layers contribute significantly to optically thick ones. This effect is larger for high-resolution spectra (top) than for the spectra binned to $\SI{0.02}{\micro m}$ (bottom). In the 10.5~$\mu$m opacity window, essentially all of the absorption occurs in the troposphere.} 
\label{fig:layer}
\end{figure}

To study the contributions of higher layers to the effective thickness, we plot in Fig. \ref{fig:layer} several effective thickness spectra for which we truncate the discretized integral (Eq. \ref{eq:integral}) at heights ranging from 6 to $\SI{126}{km}$. We do this for the high-resolution data (top panel) and for a version binned to $\SI{0.02}{\micro m}$ (bottom panel). Strong features probe all the way into the mesosphere, while low-opacity windows probe the troposphere. Greater spectral resolution accentuates the variance in opacity, allowing one to probe both deeper and higher. 

\subsection{Optical Depth Approximation} \label{sec:tau}

The optical depth along a chord of transmittance $\T(\lambda, z)$ is:
\begin{equation}
\label{eq:tau}
\tau(\lambda, z) = -\ln{\T(\lambda, z)}.
\end{equation}
An optical depth of 0 at a given wavelength implies complete transparency. A medium with an optical depth above a critical value $\tau_\mathrm{c}$ is opaque.

Effective thickness does not represent the physical thickness of the atmosphere, as layers above $h_\lambda$ may also contribute significantly to the transit depth, and layers below $h_\lambda$ may not be entirely opaque. Rather, $\Rp + h_\lambda$ corresponds to the radius of a solid body with the same transit depth as the planet.

\begin{figure}
\includegraphics[width=\columnwidth]{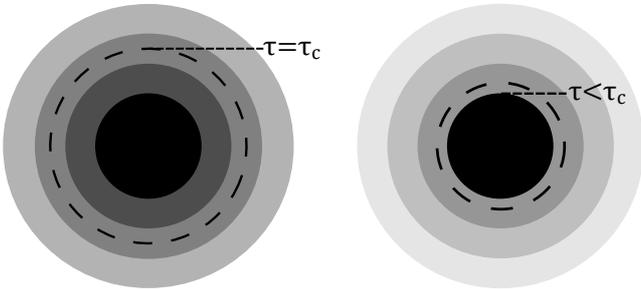}
\caption{Cartoon showing light transmitted through layers of a planet's atmosphere, at a wavelength to which the atmosphere is opaque (left) and one to which it is transparent (right). The dashed circles represent the effective thickness. In the opaque case, the effective thickness corresponds to an optical depth of $\tau_\mathrm{c} = 0.56$. In the transparent case, while $\tau < \tau_\mathrm{c}$ everywhere above the surface, the atmosphere still contributes to the transit depth, and thus the optical depth approximation is not valid.}
\label{fig:taucartoon}
\end{figure}

Nonetheless, it has been noted by previous authors that, to good approximation, the atmospheric thickness is the altitude at which the optical depth equals some critical value, $\tau_\mathrm{c}$. \citet{burrows_2007} assume $\tau_\mathrm{c} = 2/3$; \citet{lecavelier_des_etangs_rayleigh_2008} find that for many planets, $\tau_\mathrm{c} \approx 0.56$. As shown in Fig.~\ref{fig:tau}, the optical depth approximation agrees with the exact integral, except in low-opacity windows. This is because at opaque wavelengths, the contributions of different layers average out to give an effective thickness equal to that of the layer where $\tau=\tau_\mathrm{c}$. On the other hand, at transparent wavelengths, the atmosphere may never reach $\tau_\mathrm{c}$, even near the surface, as illustrated in Fig. \ref{fig:taucartoon}; hence $h_\lambda=0$ according to the optical depth approximation \citep[see also][]{betremieux_analytical_2017}. In the limit of very low opacity, the effective thickness is non-zero but also very small. We discuss the case of the sodium D-lines in Appendix~\ref{Na}.

\section{Simulating Data}\label{realistic_data}

\subsection{Refraction} \label{sec:refraction}

The effects of atmospheric refraction on a planet's transit spectrum have been extensively studied by other researchers \citep{garcia_munoz_2012, betremieux_2013, betremieux_impact_2014, betremieux_refraction_2015, misra_effects_2014, dalba_refraction_2015, betremieux_2016, robinson_analytic_2017, betremieux_hidden_2018}. We here summarize the observational effect and how we account for it in our model.  

Light travelling through the atmosphere is bent toward the planet. The deflection of rays toward the planet increases with atmospheric pressure. Refraction depends on several factors, including the range of directions of incoming starlight. \citet{betremieux_impact_2014} showed that the smaller the angular size of the host star in the planet's sky, the more significant the effect of refraction and the higher the maximum pressure (Eq. \ref{eq:pmax}). Since main sequence stars are much larger than Earth, the range of angles for incoming rays is essentially governed by the angular size of the star. In general, however, the size of the planet matters, too.

As shown by \cite{sidis_transits_2010},  \cite{betremieux_impact_2014}, and  \cite{robinson_analytic_2017}, refraction dictates that each exoplanet has a maximum pressure, $p_\mathrm{max}$, to which its atmosphere can be probed during transit:
\begin{equation} \label{eq:pmax}
\frac{p_\mathrm{max}}{p_0} = \frac{1}{\nu_0}\frac{R_*+\Rp}{a}\sqrt{\frac{H}{2\pi \Rp}},
\end{equation}
where $p_0$ is the surface pressure, $\nu_0$ is the refractivity of the atmosphere, equal to one minus the index of refraction, $a$ is the planet's orbital distance, and $H$ is the atmospheric scale height, which we take as constant. Although atmospheric refractivity is a function of wavelength, \cite{betremieux_impact_2014} note that it is essentially constant in the infrared.

The entire atmosphere can be probed if $p_\mathrm{max} > p_0$; otherwise, the minimum impact parameter, $b_\mathrm{min}$, can be determined for an exponential atmospheric pressure profile:
\begin{equation} \label{eq:bmin}
\frac{p}{p_0} = e^{-(b_\mathrm{min}-\Rp)/H}.
\end{equation}

For the Earth-Sun system, taking $\nu_0 \approx 2.9\times10^{-4}$ \citep{betremieux_refraction_2015} and $H \approx \SI{8.8}{km}$ \citep{kaltenegger_transits_2009}, we obtain $p_\mathrm{max} \approx \SI{0.24}{bar}$ and $b_\mathrm{min} \approx \SI{12.6}{km}$ \citep{garcia_munoz_2012, betremieux_impact_2014}. This means that transit spectroscopy would not be sensitive to Earth's troposphere. In particular, the 10-$\SI{12}{\micro m}$ window, where the effective thickness is as low as $\sim \SI{4}{km}$, is shifted up to a minimum effective thickness above $b_\mathrm{min}$; therefore, all detailed information from the troposphere is lost. \footnote{\textit{ACE-FTS} continues to observe the refracted image of the Sun, so there is no refraction-imposed minimum altitude in the data.} 

To account for refraction, we follow \cite{betremieux_2013} and \cite{robinson_analytic_2017}, setting transmittance to zero for all rays passing below the $b_\mathrm{min}$ of the system in question. Since the data are sampled at altitude intervals of $\SI{4}{km}$, we estimate the transmittance at $b_\mathrm{min}$ using linear interpolation between the two points straddling $b_\mathrm{min}$, and calculate the transit depth by integrating up from this point. Spectral features are compressed, but often still visible, when $b_\mathrm{min}$ is greater than their original height \citep{betremieux_impact_2014, betremieux_analytical_2017}. 

Habitable planets orbiting white dwarfs (WDs) would be relatively easy to characterize via transit spectroscopy due to the favourable planet/star radius ratio, as the signal from a planet in a transit spectrum is inversely proportional to the stellar area \citep{agol_wd_2011, fossati_wd_2012,barnes_wd_2013, loeb_wd_2013, sandhaus_wd_2016}. For comparison, a late M-dwarf's radius may be an order of magnitude smaller than the Sun's, which amplifies the signal by a factor of 100. A WD is comparable in size to Earth, so that $R_\mathrm{WD}\sim0.01R_\odot$; thus, the signal would be further amplified by a factor of $100$ compared to the M-dwarf scenario, assuming the entire WD is obstructed during transit. It is not yet known whether WDs host habitable planets.

\begin{figure*}
\includegraphics[width = \textwidth]{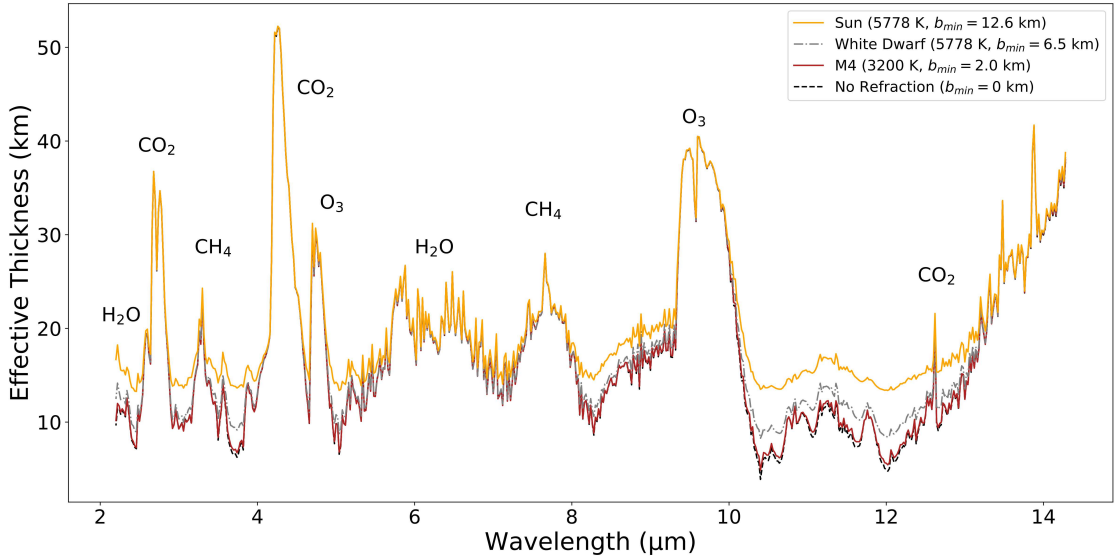}
\caption{Effective thickness spectra, binned to $\SI{0.02}{\micro m}$, for Earth analogues orbiting various stars, where we have accounted for the effects of refraction following \protect\cite{robinson_analytic_2017}. Optically thin spectral features are flattened and shifted upward, but not entirely obscured, in the corrected spectra.  
For main sequence stars, the effects of refraction are greatest for hotter stars, because their angular size is smaller as seen from a habitable planet. For Earth analogues orbiting a white dwarf, however, the planet's size is comparable to the star's, which reduces the refraction effects compared to a main-sequence star of the same temperature.} 
\label{fig:refraction}
\end{figure*}

\citet{mctier_exotopography_2018} explored the possibility of detecting topography on terrestrial planets transiting WDs, but their model did not include refraction.  Fig.~\ref{fig:refraction} shows the effect of refraction on an effective thickness spectrum for the Earth-Sun system and for two identical planets receiving the same stellar flux, one orbiting a $\SI{3200}{K}$ M-dwarf and the other a WD of radius $R_\mathrm{WD}=R_\oplus$ and temperature $T_\mathrm{WD}=T_\odot$. The effect of refraction is greater for the Sun than for an equal-temperature WD because the size of the planet is no longer negligible in the $R_*+R_p$ term in Eq.~\ref{eq:pmax}. Nonetheless, transit spectroscopy of an Earth analogue in the habitable zone of a WD cannot probe below 6.5~km---even in an opacity window and in the absence of clouds---making exotopography impossible.

\subsection{Noise} \label{sec:noise}

The primary source of noise in exoplanet transit spectra acquired from space is in the number of stellar photons received by an instrument. The number of stellar photons, $N$, in a given wavelength bin is \citep[adapted from][]{cowan_characterizing_2015}:
\begin{equation} \label{eq:noise}
N = \frac{\epsilon\Delta tA}{hc}\left(\frac{R_*}{d}\right)^2 \int _{\lambda_1}^{\lambda_2} B(\lambda,T_*)\lambda d\lambda,
\end{equation}
where $\epsilon$ is the efficiency of the instrument, $\Delta t$ is the integration time, $A$ is the collecting area of the telescope, $h$ is Planck's constant, $c$ is the speed of light, $R_*$ is the star's radius, $d$ is the distance to the system, $\lambda_1$ and $\lambda_2$ are the limits of the wavelength bin, and $B(\lambda,T_*)$ is the star's blackbody flux. In the large number Poisson limit, the fractional uncertainty on each spectral bin is $\sqrt{2/N}$, where the $\sqrt{2}$ arises from the fact that transit spectroscopy is a differential measurement (we presume equal in-transit and out-of-transit baseline). Note that $\Delta t$ refers to the total in-transit integration time, which may exceed the duration of a single transit if several transits are combined \citep[e.g.,][]{Kreidberg2014}. \citet{Lustig-Yaeger2019, Wunderlich2019} use more realistic noise models; however, eperience with the Hubble and Spitzer telescopes suggests that observers approach the photon limit shortward of $\SI{6}{\micro m}$ \citep{cowan_characterizing_2015}.

\subsection{An Earth orbiting TRAPPIST-1}
Once it is launched in the spring of 2021, the \textit{James Webb Space Telescope (JWST)} will observe Earth-like exoplanets orbiting small stars. The \textit{Near-Infrared Spectrograph} (\textit{NIRSpec}) and the \textit{Mid-Infrared Instrument} (\textit{MIRI}) together cover the entire wavelength range of our empirical Earth spectrum \citep{gardner_jwst_2006}.

\begin{figure}
\includegraphics[width=\columnwidth]{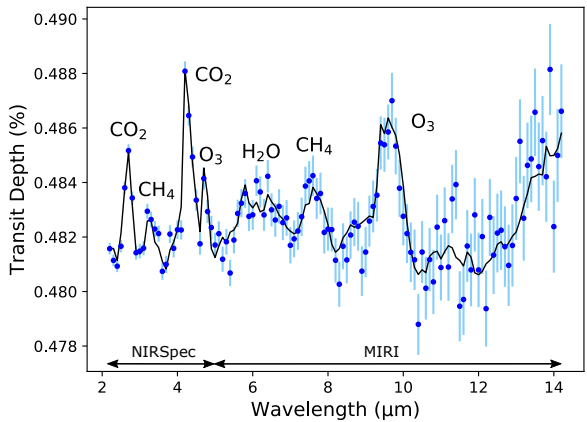}
\caption{Mock \textit{JWST} spectrum for TRAPPIST-1e, using \textit{NIRSpec} and \textit{MIRI}. The black line is the model, made by adapting Earth's spectrum to the appropriate planetary and stellar radii and binning it to $\SI{0.1}{\micro m}$. Mock data are normally distributed around the model. Error bars correspond to photon noise from the star, assuming 5.8 days of integration time (approximately 2 weeks of total observing time), or about 150 transits with each \textit{NIRSpec} and \textit{MIRI} over the course of \textit{JWST}'s 10 year  expected lifetime. In the near-infrared, CO$_2$, O$_3$, and CH$_4$ are detectable at levels above $40\sigma$, $10\sigma$, and $7\sigma$, respectively. In the mid-infrared, O$_3,  $ H$_2$O, and CH$_4$ can be detected at around $10\sigma$, 6$\sigma$, and $5\sigma$, respectively. Note that as the significance scales as the square root of the number of transits, it would take far fewer transits to detect most of these features at the 5$\sigma$ level, as discussed in Section~\ref{sec:discussion}. Since \textit{NIRSpec} and \textit{MIRI} cannot be used in parallel, obtaining the full spectrum at this signal-to-noise would require more than \textit{JWST}'s 10 year mission duration. Short-wavelength spectral features are easier to detect due to falling stellar flux at longer wavelengths. Nonetheless, the biosignatures CH$_4$ and O$_3$ could in principle be detected with either \textit{NIRSpec} or \textit{MIRI}.}
\label{fig:jwst}
\end{figure}

We use the \textit{ACE-FTS} data to build a mock \textit{JWST} transit spectrum of an exo-Earth. The TRAPPIST-1 system is promising, as several of its planets are Earth-sized and in the habitable zone, the star is small, and the system is nearby \citep{gillon_trappist_2016, gillon_trappist_2017, dewit_2016, Luger_2017, delrez_trappist_2018}. Fig. \ref{fig:jwst} shows a mock transit spectrum for TRAPPIST-1e. We use the planet's measured radius \citep{delrez_trappist_2018}, but assume for illustrative purposes that its atmospheric composition and scale height are identical to Earth's; this is unlikely to be the case \citep{Grimm_2018} but is consistent with current observational constraints \citep{deWit2018}. Following \citet{deming_2009}, we assume that every observable transit in a ten-year mission is observed; this corresponds to about 5.8 days of in-transit integration time, or 150 transits. Since \textit{NIRSpec} and \textit{MIRI} cannot be used simultaneously, it would take longer than \textit{JWST}'s entire nominal mission lifetime to acquire the entire spectrum at this SNR. 

As shown in Fig.~\ref{fig:jwst}, a dedicated \textit{JWST}/NIRSpec campaign could robustly detect CO$_2$, while H$_2$O can be detected at lower significance with \textit{NIRSpec}, and  O$_3$ and CH$_4$, with \textit{NIRSpec} or \textit{MIRI}. While the latter two molecules produce larger-amplitude spectral features in the mid-infrared, a \textit{MIRI} campaign would be more challenging due to lower stellar flux and increased thermal noise. 

\section{Discussion: Habitability \& Biosignatures} \label{sec:discussion}
It is notoriously difficult to measure abundances from transit spectra \citep[e.g.,][] {burrows_2014}. Following \cite{deming_2009}, we conservatively estimate the detection significance of a molecule as proportional to the ratio of the amplitude of the dominant spectral feature to its uncertainty. This quantity scales as the square root of the number of data points in the feature. We find that in the near-infrared, a 5$\sigma$ detection of CO$_2$ could easily be achieved in \textit{JWST}'s lifetime: this would require as few as 3 transits. It would take closer to 40 and 80 transits to detect O$_3$ and CH$_4$ at this significance in the near-infrared; in the mid-infrared, these latter two molecules could be detected in approximately 80 and 150 transits, respectively, while H$_2$O could be detected in 125 transits.

As TRAPPIST-1e can be observed for approximately 81 transits in the 5 years of JWST's nominal lifetime \citep{Lustig-Yaeger2019}, our results suggest that the biosignatures would be difficult to detect on this planet. However, using simulations of Earth with M-dwarf host stars, \citet{Wunderlich2019} find that the TRAPPIST-1 planets could have approximately an order of magnitude less O$_3$ and three orders of magnitude more CH$_4$ than Earth; this would make the CH$_4$ features much easier to detect, while slightly compromising the detectability of O$_3$.

Moreover, the detectability of a spectral feature depends on many factors, such as the bulk composition of the atmosphere and its scale height. \citet{Lustig-Yaeger2019} assess the detectability of spectral features on the TRAPPIST-1 planets for several possible atmospheric compositions. Although they do not consider Earth's 1 bar N$_2$/O$_2$ atmosphere, they find that CO$_2$ -- the most readily detectable molecule in Earth's spectrum -- consistently produces the largest feature, whose amplitude is insensitive to the planet's simulated evolution or atmospheric composition. This feature can therefore be used to establish the presence of an atmosphere, which can be further characterized using other molecular signatures. They find that depending on the scenario, it would take between 2 and 30 transits to show that a planet's spectrum is not flat, consistent with our findings.

The greatest threat to the habitability of a terrestrial planet in the habitable zone of its host star is the lack of an atmosphere, and the lack of water. For the purposes of establishing the habitability of such a planet, it is necessary, and arguably sufficient, to show that the planet has an atmosphere with water. The detection of carbon dioxide and water vapour in the planet's transit spectrum would not prove that habitable conditions exist at its surface, but would be highly suggestive. Likewise, the mere presence of detectable quantities of methane and ozone is a biosignature \citep{sagan_biosignatures_1993}. For more nuanced perspectives on establishing the habitability and inhabitation of exoplanets, see \citep{Robinson2017,Schwieterman2018}.

It is worth comparing our empirical transit spectrum and its biologically-sourced features to simulations performed by previous authors. \cite{kaltenegger_transits_2009} and \cite{Barstow2016} are the only previous papers to have considered the transit spectrum of Earth including biosignatures.

\cite{kaltenegger_transits_2009} presented simulated spectra, but their model had been calibrated against Solar occultation spectra similar to the \textit{ACE-FTS} data we have used.  They did not specifically consider the TRAPPIST-1 planets, as they had not yet been discovered.  Extrapolating from their Table~2, their results suggest that a 300-transit campaign with \textit{MIRI} would detect the $\SI{9.9}{\micro m}$ ozone feature at $13\sigma$ and the $\SI{7.5}{\micro m}$ methane feature at about $4\sigma$, in broad agreement with our results.

\cite{Barstow2016} simulated transit spectra of TRAPPIST-1b, c, and d, assuming Earth-like atmospheric compositions, assessing the detectability of ozone with \textit{JWST/MIRI}. They find that both the $\SI{4.5}{\micro m}$ and the $\SI{9}{\micro m}$ ozone features of the \cite{gillon_trappist_2016} TRAPPIST-1d would be detectable by combining data from 30 transits. But \cite{gillon_trappist_2017}  discovered new planets in the system and completely revised the parameters of the outer planets, including d.  As result, the \cite{Barstow2016} signal to noise estimates for that planet are no longer accurate.

Finally, many authors have considered transit spectra of ``Earth-like'' planets, but without modern Earth's biosignatures.  \cite{deming_2009} simulated \textit{NIRSpec} transit spectra for an Earth-like planet, but only considered water vapour and carbon dioxide, so could not evaluate the detectability of atmospheric biosignatures. \cite{Morley2017} simulated TRAPPIST-1e as an \emph{abiotic} Earth and hence their transmission spectra show an extremely weak methane feature at 7-$\SI{8}{\micro m}$, and no ozone feature at $\SI{9}{\micro m}$. They did not discuss the detectability of atmospheric biosignatures, since there were none in their simulations. \cite{Batalha2018} likewise simulated an Earth-like TRAPPIST-1f with a variety of abiotic atmospheric compositions: they have no simulation including oxygen and/or ozone and hence could not comment on the detectability of those features. Moreover, they consider emission, rather than transit, spectroscopy with \textit{MIRI}. \cite{2018Krissansen-Totton} studied the detectability of biosignatures in the simulated \textit{NIRSpec} transit spectra of an \emph{anoxic} TRAPPIST-1e, so methane was detectable but oxygen and ozone were non-existent.

\section{Summary} \label{sec:conclusion}
We have used Solar occultations of Earth's atmosphere from \textit{ACE-FTS} to build an empirical infrared transit spectrum of Earth to quantify the extent to which we can probe the atmospheres of Earth analogues orbiting nearby stars. 

Even though the effective thickness of Earth's atmosphere is greater than a few km at all wavelengths considered, the surface transmittance in opacity windows can be order unity. Unsurprisingly, the maximum surface transmittance increases with spectral resolution.  This suggests that one could---in principle---probe the surface environment of Earth analogues via high-resolution transit spectroscopy.  

However, we used the cloud-free \textit{ACE-FTS} data to construct our transit spectra, so our results represent the deepest possible look into Earth's atmosphere. In general, clouds and aerosols are present in the troposphere and sometimes in the stratosphere (\citealt{junge_clouds_1961}; see \citealt{wang_clouds_2013} for a comprehensive description of clouds). Clouds increase the opacity of a planet's lower atmosphere, partially obscuring spectral features \citep{howe_2012, benneke_2012, benneke_2013, betremieux_impact_2014, betremieux_analytical_2017}.

Moreover, atmospheric refraction limits the depth to which we can probe planets in transit.  While this effect is small for Earth analogues orbiting M-dwarfs, it is considerable for Sun-like stars and white dwarfs. 

Lastly, a multi-year campaign with the \textit{James Webb Space Telescope} targeting an Earth-analogue orbiting an M-dwarf could detect carbon dioxide in the near-infrared, water vapour in the mid-infrared, and the biosignature duo of methane and ozone in the near or mid-infrared. In particular, the detection of ozone and methane on TRAPPIST-1e with \textit{JWST} would be difficult---but not impossible.  A larger collecting area, such as the Origins Space Telescope \citep{OST2018}, would help. The Transiting Exoplanet Survey Satellite will be able to search for systems with younger M-dwarf host stars in \textit{JWST}'s continuous viewing zone \citep{deming_2009, Ricker2015}. Despite the larger number of observable transits, the earlier spectral types may mean that such systems are not easier to study with \textit{JWST}. \textit{SPECULOOS} \citep{Burdanov2018} is designed to search for potentially habitable planets transiting nearby ultracool dwarfs, and is therefore better suited to find candidates for atmospheric characterization. The \textit{MEarth} transit survey \citep{Nutzman2008, Irwin2009} will be able to find habitable zone super Earths around late M-dwarfs. High-resolution ground-based spectrographs such as \textit{CARMENES} \citep{Quirrenbach2014} and \textit{SPIRou} \citep{Donati2018} will be able to discover M-dwarf planets through radial velocity observations.

The co-existence of remotely-detectable methane and ozone is thought to be a relatively recent development on Earth \citep{Lyons2014,Schwieterman2018,Reinhard2017, Bellefroid2018,Crockford2018}, so this search is both high-risk and high-reward.

\section*{Acknowledgements}

E.J.R.M. is funded by a McGill Science Undergraduate Research Award. The authors thank Yi Huang for his insight on the \textit{ACE-FTS} data and on stratospheric clouds, Mark Swain for the idea of investigating Earth's sodium layer, and Emily Pass for her helpful comments on this work. We also thank Yan B\'etr\'emieux, Tyler Robinson, and an anonymous referee for their comments, which significantly improved this work. N.B.C. acknowledges funding from NSERC and FRQNT.

\bibliographystyle{mnras}

\appendix

\section{Sodium D-Lines of Earth}\label{Na}

Stellar occultation measurements from \textit{GOMOS} yield the vertical dependence of the slant transmission of Earth's atmosphere at wavelengths including the sodium D-lines \citep{Fussen2004}. Since sodium has already been detected in the atmospheres of exoplanets via transit spectroscopy from space and the ground \citep[e.g.,][]{Charbonneau2002,Redfield2008}, it is worth asking whether Earth's atmospheric sodium is remotely detectable.

The vertical distribution of Na in Earth's atmosphere leads to roughly constant transmittance ($1-\T \approx 10^{-3}$ at $R=2000$) over the range $b\in[0,90]$~km. We now derive an approximate expression for the atmosphere's effective thickness that is valid in this case.

The second term under the square root in Eq.~\ref{h_eqn} is much smaller than the first if the transmittance is close to unity and/or the vertical extent of the absorbing layer is small compared to the planetary radius. Since both conditions hold for Earth's sodium layer, we Taylor expand the square root to obtain
\begin{equation}
h_\lambda \approx \frac{1}{R_p} \int_{\Rp}^{R_*} b(1-\T(b,\lambda))db.
\end{equation}

If we presume the transmittance to be uniform up to some impact parameter $b=\Rp+z_0$, and zero elsewhere, then the integral simplifies to
\begin{equation}
h_\lambda \approx \frac{1-\T(\lambda)}{R_p} \int_{\Rp}^{\Rp+z_0} bdb = \frac{1-\T(\lambda)}{2 R_p} \left(2\Rp z_0 + z_0^2\right).
\end{equation}

Now if we further presume that $z_0 \ll \Rp$, then
\begin{equation}
h_\lambda \approx \big(1-\T(\lambda)\big) z_0.
\end{equation}
Therefore, the effective thickness of Earth's atmosphere in the sodium D-layer is only 90~metres and therefore undetectable at $R=2000$.

\bsp	
\label{lastpage}
\end{document}